%% file: main-arxiv.tex
\documentclass[a4paper,11pt]{article}
\pdfoutput=1
\usepackage{jinstpub} 
\usepackage{lineno}
\usepackage[nolist]{acronym}
\usepackage{subcaption}
\usepackage{tikzit}
\input{sample.tikzstyles}

\title{Surface and Near-Surface Positron Annihilation Spectroscopy at Very Low Positron Energy}

\author[a,b]{Lucian Mathes,}
\author[a]{Maximilian Suhr,}
\author[b]{Vassily V. Burwitz,}
\author[a]{Danny R. Russell,}
\author[a]{Sebastian Vohburger,}
\author[a]{and Christoph Hugenschmidt}
\affiliation[a]{Heinz Maier-Leibnitz Zentrum (MLZ), Technical University of Munich,\\Lichtenbergstr. 1, 85748 Garching, Germany}
\affiliation[b]{Technical University of Munich, School of Natural Sciences, Physics Department,\\ James-Franck-Str. 1, 85748 Garching, Germany}

\emailAdd{lucian.mathes@tum.de}

\abstract{
We present a monoenergetic positron beam specifically tailored to the needs of (near-) surface positron annihilation spectroscopy. The \acf{slope} comprises a high-activity $^{22}$Na source, a tungsten moderator, electrostatic extraction and acceleration, magnetic beam guidance, as well as an analysis chamber with a movable sample holder and a $\upgamma$-ray detection system.
The tungsten moderator foil, biased between 0 and 30~V, in combination with the HV-biasable sample holder, enables positron implantation energies between 3~eV and 40~keV.
At low energies ($<$20~eV), the count rate typically amounts to 4400~counts per second, and the beam diameter is smaller than $12\pm3$~mm.
We conduct phase space simulations of the positron beam using COMSOL Multiphysics$^\circledR$ to characterize the beam properties and compare the findings with the experimentally determined energy-dependent beam diameter.
To showcase the capabilities of \ac{slope}, we perform studies of positronium (Ps) formation on boehmite and depth-resolved \acf{cdbs} of copper.
In particular, the Ps formation at the hydrogen-terminated surface of boehmite is found to be maximum at a positron implantation energy of 10~eV.
The range of positron energies for which we observe Ps formation agrees with the hydrogen ionization energy.
}

\begin{document}
\maketitle
\flushbottom

\begin{acronym}
\acro{slope}[SLOPE]{Setup for LOw-energy Positron Experiments}
\acro{v2p}[V/P]{valley-to-peak}
\acro{fwhm}[FWHM]{Full Width at Half Maximum}
\acro{tmp}[TMP]{turbomolecular pump}
\acroplural{tmp}[TMPs]{turbomolecular pumps}
\acro{uhv}[UHV]{ultra-high vacuum}
\acro{pas}[PAS]{Positron annihilation spectroscopy}
\acro{dbs}[DBS]{Doppler-broadening spectroscopy}
\acro{cdbs}[CDBS]{coincidence Doppler-broadening spectroscopy}
\acro{hpge}[HPGe]{high-purity Ge}

\end{acronym}

\section{Introduction}\label{sec:intro}
The investigation of materials surfaces and interfaces is of utmost interest in understanding their properties, which are crucial for many scientific and technological applications.
\ac{pas} is a powerful non-destructive tool for probing defects at the atomic scale (see, e.g., \cite{rehberg1999, cizek2018}).
In particular, \ac{pas} using a monoenergetic positron beam provides valuable insights into the electronic structure, defect characteristics, and chemistry of the (near-) surface region of materials \cite{uedono2003, pikart2011, Hug16}.
For fundamental research, knowledge of surface properties is crucial for designing surfaces tailored specifically for the efficient formation of ortho-positronium (o-Ps) \cite{mills1978, cassidy2007, Gua22} needed for the production of anti-hydrogen \cite{brusa2017} or the emission of positronium negative ions Ps$^-$ \cite{Cee11b, nagashima2014}.
To further advance our understanding of surface physics and chemistry, cutting-edge positron beam setups are required.

We introduce the unique \acf{slope}, which is particularly tailored to the needs of (near-) surface PAS with a monoenergetic positron beam.
\ac{slope} is designed to allow for precise and detailed investigations of surface properties, positron-surface interactions, and the formation and emission of Ps.
In addition, \ac{slope} also allows bulk and depth-dependent studies with positron implantation energies up to 40~keV.
\ac{slope} is equipped with a pair of \ac{hpge} detectors to enable (i) depth dependent \acf{dbs} for the determination of positron diffusion lengths and defect distributions, (ii) \acf{cdbs} for investigating the chemical surrounding of vacancy-like defects and the identification of precipitates, and (iii) the observation of free o-Ps via the detection of 3$\upgamma$-events with high energy resolution. 

This paper provides a comprehensive overview of \ac{slope}, detailing its design and operational capabilities and highlighting some potential applications.
We describe the key components of the setup, including the positron source and moderation assembly, electromagnetic beam guidance system, sample chamber, and detection system.
Additionally, we compare measurement results to simulations for the determination of beam characteristics such as diameter and transverse momentum spread.
To showcase the versatility and performance of \ac{slope}, we compare depth-dependent \ac{dbs} and \ac{cdbs} of Cu. 
Finally, we present an experiment correlating the kinetic energy of slow positrons with the Ps formation fraction on the hydrogen-terminated surface of boehmite. 


\section{Experimental Setup}\label{sec:setup}
\begin{figure}[htbp]
\centering
\begin{subfigure}{\textwidth}
\resizebox{\textwidth}{!}{
  \input{figures/labbeam.tikz}
}
\end{subfigure}\\
\vspace{5mm}
\begin{subfigure}{\textwidth}
\includegraphics[trim={10mm 15mm 10mm 15mm},clip,width=\textwidth]{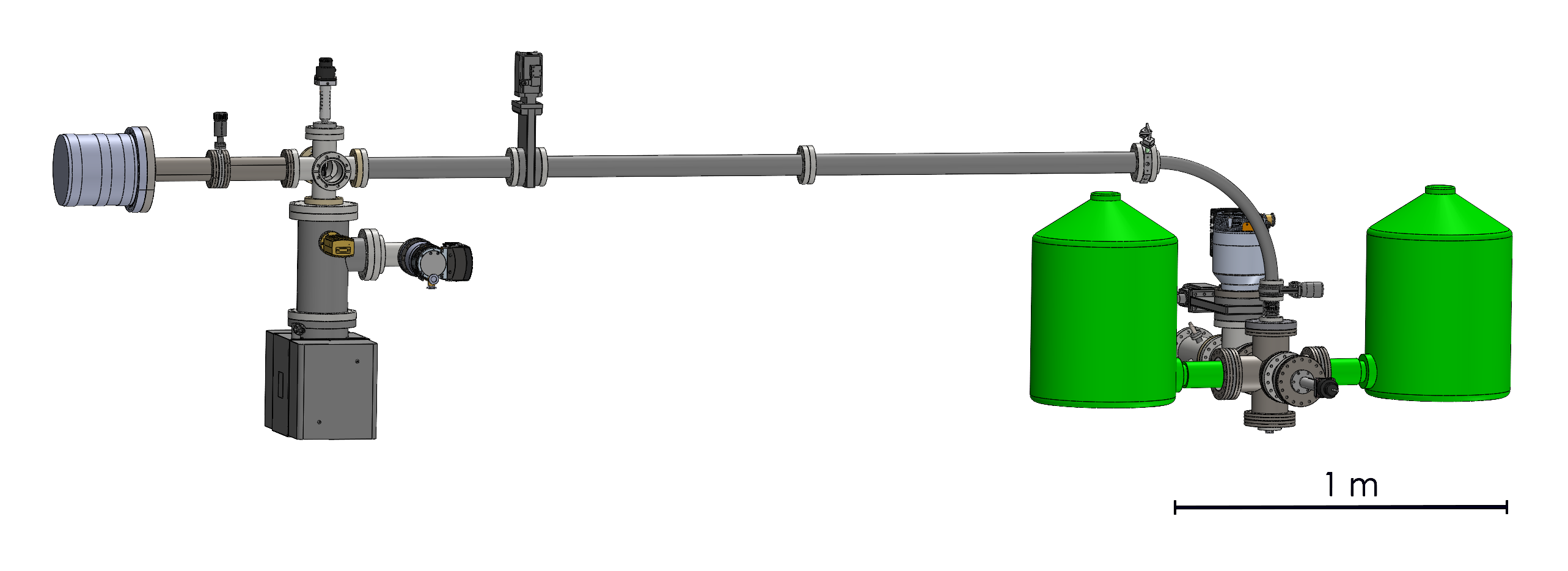}
\end{subfigure}

\caption{\label{fig:setup}
Schematic (top) and CAD (lower) drawing of SLOPE.
LA, LB, and LC denote the cylindrical acceleration and beam formation electrodes.
Solenoids, Helmholtz coil pairs, and flange coils for magnetic beam guidance are labeled G, H, and F, respectively.
Transverse magnetic fields are applied by saddle coils (CH and CV) to compensate the earth's magnetic field and the curvature drift.
(Lead shielding not shown.)
}
\end{figure}

\begin{figure}[htbp]
\centering
\includegraphics[width=\textwidth]{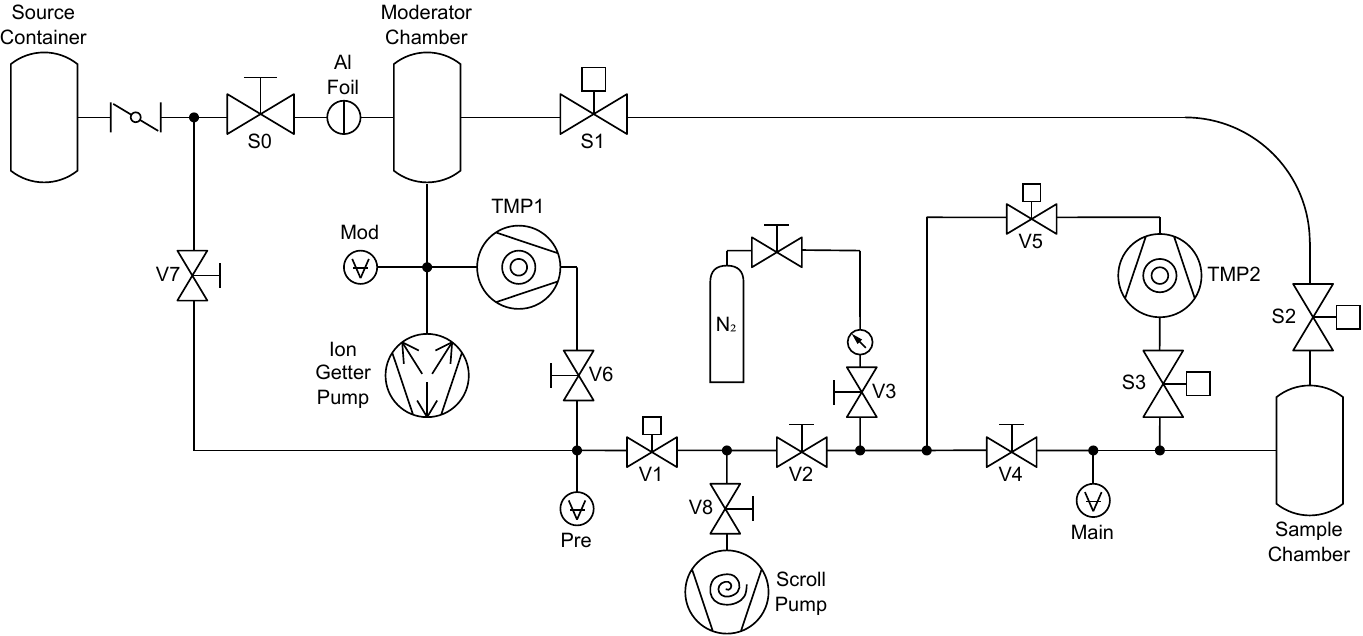}
\caption{\label{fig:vacuum}
Schematic of the vacuum system.
Pre-vacuum valves and high-vacuum shutters are labeled V1 to 8 and S0 to 3, respectively.
Pneumatic valves/shutters (in contrast to manual ones) are marked with an empty square handle.
The $10~\upmu$m Al foil separates the pre-vacuum section on the left from the \ac{uhv} of the moderator and sample chamber.
Both \acfp{tmp} are connected to the same scroll pump.}
\end{figure}

An overview of \ac{slope} is shown in Figure~\ref{fig:setup}, and a schematic of the vacuum system is plotted in Figure~\ref{fig:vacuum}.
The present setup results from an upgrade of the positron beam dedicated to positron-induced Auger-electron spectroscopy (PAES) \cite{strasser1999, strasser2001}. 
The source container and the moderator chamber remained unchanged, whereas most parts of the beamline, the sample chamber with movable sample holder, and the detection system were newly built.

\subsection{Vacuum System}\label{sec:vacuum}
The vacuum system is separated into three main sections:
1) The positron source with shielding container and transfer rod, 2) the beamline with moderator and electric lenses, and 3) the sample chamber.
A scroll pump (Edwards ESDP12) is directly connected to the source container, which is separated from the rest of the setup by a $10~\upmu$m  Al foil.
The scroll pump also constitutes the pre-vacuum for two \acp{tmp} located at the moderator chamber and the sample chamber (both Oerlikon Leybold).
Both pumps can reach ultra-high vacuum ($\sim10^{-8}~\text{mbar}$) within a few hours.
An additional ion getter (Ti sublimation) pump (Varian StarCell) is located at the moderator chamber.
Attached to each chamber is a hot cathode ionization pressure gauge (Leybold Ionivac ITR 200 S).
The two chambers can be separated from the beamline using electro-pneumatic gate valves.
Before changing the sample, the sample chamber is flooded with N$_2$.

\subsection{Positron Source and Moderator Assembly}\label{sec:source}
The $^{22}$Na source (purchased from iThemba LABS) has an activity of 1.16~GBq (as of June 2024).
The $^{22}$Na is sealed in a standard capsule with a Ta reflector at the back and a 5~$\upmu$m thick Ti window with a diameter of 4~mm at the front \cite{rehberg2004}.
The source capsule is mounted on a transfer rod, which can be positioned close to the moderator foil in the center of the moderator chamber.
The source container (and half of the moderator chamber) is separated from the \ac{uhv} system by a 10~$\upmu$m Al foil.
This design was chosen to preserve the rest of the \ac{uhv} system from possible contamination with radio-nuclides, e.g., in case non-encapsulated positron emitters are used, such as $^{64}$Cu \cite{strasser1999}.

Positrons emitted from the $^{22}$Na source are moderated in a thin W foil in transmission geometry.
The monocrystalline W(100) foil (purchased from the Thin Film Laboratory at Aarhus University) with dimensions 10 x 12~mm$^2$ and 1~$\upmu$m thickness is mounted in a Mo frame in a way that allows heating with an electric current \cite{strasserPHD}.
For the desorption of possible adsorbates at the surface and annealing of the W foil by current heating, the whole Mo frame is mounted on a linear feed-through.
This feed-through can be moved to an off-axis position in order to minimize the heating of the Al foil during moderator conditioning.

In normal operation mode, the moderator chamber is constantly held at pressures $<10^{-7}~\text{mbar}$ and a positive bias of typically $U_\text{mod}=30~\text{V}$ is applied to the moderator for positron extraction.
Combined with the negative positron work function for W(100), $\Phi_+^\text{W}=-3.0~\text{eV}$ \cite{hugenschmidt2002}, this results in a beam energy of $E=-\Phi_+^\text{W}+U_\text{mod}=33~\text{eV}$.

\subsection{Beam Guidance System}\label{sec:beam-guidance}

The whole positron beam guidance system consists of 29 magnetic coils, 3 beam formation electrodes, the (biased) moderator, and the sample holder ($0$ to $-40~\text{kV}$).
All coils and potentials can be remotely controlled, enabling quick automatic optimization and variation of the beam guiding parameters at all energies.
The 29 magnetic coils (see labeling in Figure \ref{fig:setup}) are:
\begin{itemize}
		\item 4 guidance solenoids (G1-4)
		\item 3 Helmholtz coil pairs (H1-3)
		\item 9 flange coils (F1-9)
		\item 4 pairs of correction (saddle) coils, horizontal/vertical (C1-4, H \& V)
\end{itemize}

The solenoids G1-4 are made of Cu wires (2~mm diameter, two layers) directly mounted on the stainless steel beam tubes. 
The solenoid coils are powered with up to 8~A to generate the longitudinal magnetic guiding field.
For 8~A and a typical winding per unit length $N/l=1000~\text{m}^{-1}$ the field strength is $B\approx10~\text{mT}$.
Helmholtz-like pairs of coils are used instead of solenoid coils, where the latter cannot be installed due to geometrical constraints.
The coils are mounted around cross-piece flanges and operated with currents of about 5~A.
One Helmholtz pair is used at the moderator chamber, where a homogeneous field is particularly beneficial for positron extraction.
Two Helmholtz pairs are attached to the sample chamber.
In addition, flange coils ($\approx5~\text{A}$) are installed to increase the homogeneity of the guiding field at gaps between the solenoids.

Two pairs of long saddle coils with typically 2 to 4 windings are glued onto each solenoid to create transverse magnetic fields.
They are operated with currents of order 1~A to produce a field in the horizontal ($\vec{B}_H$) and vertical ($\vec{B}_V$) directions with respect to the longitudinal guiding field.
This compensates for external magnetic fields and gradient/curvature drift in bends.

Electrostatic acceleration and beam formation are realized by potentials applied to the moderator, three electrodes, and the biasable sample holder.
The electrodes LA, LB and LC are used for positron extraction and beam alignment \cite{strasserPHD}.
The moderator is typically operated at $U_\text{mod}=30~\text{V}$ but can be set to any lower value if particularly low positron kinetic energy is required.
The sample potential can be varied from $0$ to $-40~\text{kV}$ via a high-voltage power supply unit (FUG HCP 35-65000).
Consequently, \ac{slope} covers a large range of positron implantation energies from 3~eV to 40~keV.


The beamline is equipped with an optional aperture in a custom-made CF63 flange mounted directly upstream of the bend to facilitate beam guidance.
Without breaking the vacuum, it can be varied between four states:
Completely closed, 4~mm diameter aperture, 5~mm diameter target, or entirely open.
We use the aperture to center the beam in the first section of the beamline.
In normal operation, the aperture flange is opened entirely and does not interfere with the positron beam.

\subsection{Sample Chamber}\label{sec:sample-chamber}
The sample chamber consists of a sixfold CF100 cross-piece.
Two detector ``cups'' made of stainless steel with a wall thickness of 1.5~mm  are installed in opposing flanges, allowing two \ac{hpge} detectors to be placed close to the sample (see section \ref{sec:data-acquisition}).
Figure\,\ref{fig:trajectories} shows a cross-section through the center of the sample chamber.

The sample holder consists of an Al plate mounted to a motorized linear feed-through installed at the front of the chamber.
With a usable length of 80~mm, the sample holder provides enough space for up to five samples.
The sample holder is connected to an HV feed-through from below via an Al socket surrounded by glass-ceramics (Macor).
On top of the Al socket is a Fe pin mounted on a spring bearing, which ensures electrical contact while keeping friction to a minimum.
The Fe pin (relative permeability of about 4000) increases the local magnetic field at the sample position and hence focuses the adiabatically guided low-energy positron beam (see Figures~\ref{fig:trajectories} and \ref{fig:fwhm-vs-energy}).

\subsection{Detection System and Data Analysis}\label{sec:data-acquisition}
The annihilation radiation emitted from the sample is detected by two HPGe detectors (Ortec GEM35P4-70, efficiency $\approx35\%$).
They are placed face-to-face on individual racks, separate from the rest of the setup, to eliminate vibrations.
Both \ac{hpge} crystals of the detectors have a diameter of 58.9~mm and a length of 68~mm.
They are placed at a 45~mm distance from the center of the sample chamber.
Taking an effective detection circular area with a diameter of 50~mm results in a solid angle $\Omega_\text{det}=0.97$~sr and hence an effective field of view $\Delta\Omega_\text{det}=\Omega_\text{det}/4\pi=0.077$ per detector.
The preamplified signals of the detectors are read out via a CAEN DT5781 multi-channel analyzer.
The recorded spectra and additional metadata are saved in the N42 nuclear radiation data file format \cite{nist_n42-2020}.
The detectors can be used in coincidence mode, i.e., for \ac{cdbs}, by evaluating the event timestamps in list-mode data.
We regularly perform energy calibrations with a $^{152}$Eu source;
the determined energy resolution is typically $\approx1.5$~keV at 511~keV.

For the evaluation and analysis of the recorded $\gamma$ spectra, we use the software suite STACS \cite{chryssos2023}.
In \ac{dbs}, the line shape parameter S is typically employed to characterize the Doppler-broadened 511~keV annihilation line.
For all data presented in the following, the S parameter is defined by the ratio of counts in the energy range 510 to 512~keV and the entire (background-corrected) annihilation photo peak (491 to 531~keV).
The \ac{v2p} ratio is commonly used to describe the fraction of 3$\gamma$ annihilation events, i.e., the fraction of positrons annihilating as ortho-Positronium (o-Ps).
The \ac{v2p} value is calculated by the ratio of events in the energy range 450 to 490~keV and the uncorrected photo peak 491 to 531~keV.

\subsection{Predicted and Measured Annihilation Event Rate in the Sample}
\label{sec:intensity}
The expected number of positrons per second $I_\text{e$^+$}$ arriving at the sample can be calculated by 
\begin{equation}
I_\text{e$^+$} = A \cdot R \cdot \Delta\Omega_\text{mod} \cdot \eta_\text{Al}^{\text{e}^+} \cdot \epsilon_\text{mod},
\end{equation}
with the source activity $A=1.16$~GBq,
the $\upbeta^+$ branching ratio $R=90.3\%$ of $^{22}$Na \cite{TabRad_v5},
the effective solid angle of the moderator area with respect to the source $\Delta\Omega_\text{mod}=0.25$,
and the transmission probability of positrons through the $10~\upmu$m Al foil $\eta_\text{Al}^{\text{e}^+}=89\%$ (the average angle of positrons passing through the Al foil of $\approx30^{\circ}$ results in a mean effective thickness of $11.5~\upmu$m).
The moderator efficiency of a 1~$\upmu$m thin annealed W(100) foil is expected to be $\epsilon_\text{mod}\approx0.05\%$.
Using these values and assuming no transport loss, we get a beam intensity of about $1.2\times10^5$~moderated positrons per second at the sample position.

We further calculate the expected total count rate in each detector via
\begin{equation}
I_\upgamma = I_\text{e$^+$} \cdot (1 - \eta_\text{back}) \cdot 2 \cdot \eta_\text{wall} \cdot \Delta\Omega_\text{det} \cdot \epsilon_\text{det},
\end{equation}
with the backscattering coefficient $\eta_\text{back}=0.25$ for 30~keV positrons on Cu \cite{coleman1992} and
the factor 2 for two produced $\gamma$ quanta per positron.
Plugging in the transmission probability of 511~keV $\upgamma$'s through the chamber wall (1.5~mm of 304L stainless steel) $\eta_\text{wall}\approx87\%$ (the average angle of photons passing through the wall of $\approx45^{\circ}$ results in a mean effective  thickness of $2.1~$mm),
the field of view of one detector $\Delta\Omega_\text{det}=7.7\%$,
and the detector efficiency $\epsilon_\text{det}=35\%$,
we get a total count rate of about 4.1~kcts/s.

In experiments with Cu and 30~keV positrons, the measured total count rate is typically 3.9~kcts/s in the recorded gamma spectrum, agreeing with the estimated count rate.
Not included in this number are pile-up events filtered by the data processing algorithm; they amount to 12\% of the recorded valid events.
The so-called peak-to-total ratio, i.e., the fraction of counts collected in the 511~keV photo peak, is 29\%.

\section{Simulations}
We simulated the magnetic and electric fields of \ac{slope} and positron trajectories with COMSOL Multiphysics$^\circledR$ to understand the beam characteristics as a function of coil currents and applied potentials.
For this, we used the Electrostatics (es), Magnetic Fields (mf), and Charged Particle Tracing (cpt) modules.
The beam is modeled with positrons emitted from the moderator foil with a longitudinal energy of 3~eV, equal to the modulus of the positron work function of W, $\Phi_+^\text{W}$.
As in the experiment, the moderator potential can be varied, which -- together with the positron work function -- determines the longitudinal momentum of the positrons.
The transverse momentum of the particles is assumed to be thermally (Maxwell-Boltzmann) distributed with a mean energy of 40~meV.
We simulated the beam with 10,000 (50,000 for plots of the phase space volume) individual particles being emitted simultaneously, neglecting particle-particle interactions due to the low space charge density of the beam. 
Like in real experiments, the optimized magnetic guiding fields differ for the two moderator voltages used, the main difference being the solenoids G1-3 powered with 2~A (for $U_\text{mod}=0~$V) and 8~A (for $U_\text{mod}=30~$V), respectively.

For the spatial distribution of positrons emitted from the moderator, we consider the intensity distribution of a source irradiating a flat surface from a given distance.
At \ac{slope}, the $^{22}$Na  is presumably uniformly distributed within a diameter of 4~mm, and the source is located at a distance of 4~mm from the moderator.
The resulting distribution has a diameter of 5.8~mm \ac{fwhm}.
The area of the moderator of $10\times10$~mm$^2$ cuts the beam into a squared shape.
(We use a rather conservative value (4~mm) for the source diameter and neglect the (defocusing) scattering effects of the Ta reflector, the 5~$\upmu$m Ti window of the source capsule and the 10~$\upmu$m Al foil in front of it.)

\begin{figure}[htbp]
\centering
\includegraphics[width=.5\textwidth]{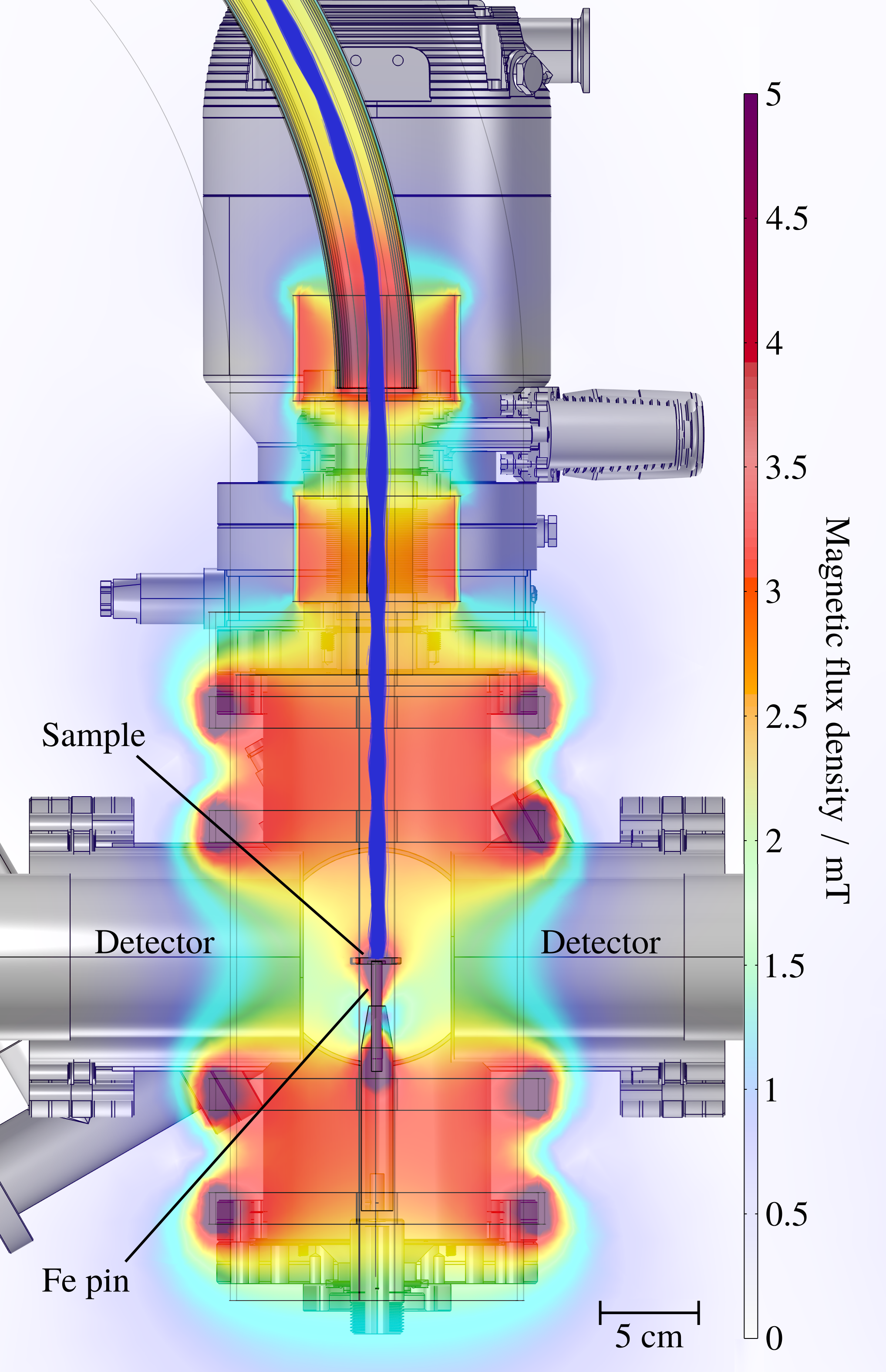}
\caption{\label{fig:trajectories}
Simulation of the magnetic field and trajectories of 3 eV positrons hitting a sample.
The moderator and sample holder are at ground potential.
The beam energy of 3~eV is hence defined by the positron work function of W. 
Only positrons emitted from the moderator within the \ac{fwhm} of 5.8~mm are shown.
The transverse velocities are assumed to be thermally distributed with an energy of 40~meV.
The Fe pin below the sample in the center of the sample chamber locally increases the magnetic field strength and thus leads to adiabatic compression of the beam.
The two \ac{hpge} detectors are placed to the left and right of the sample position.
}
\end{figure}

We have successfully used the simulations to find coil settings for a more homogeneous magnetic guidance field and to increase the count rate at the sample position at very low implantation energies $<33~$eV, i.e., for $U_\text{mod}<30~$V.
These settings produce an average magnetic flux density of 2.5~mT.
Figure~\ref{fig:trajectories} shows the simulated 3~eV beam ($U_\text{mod}=0~$V) and the magnetic flux density at the sample chamber.
For visibility, the positron distribution at the moderator was cut to a circle with a diameter equal to the 5.8~mm \ac{fwhm} of the distribution.
Note the increasing magnetic field strength above and at the sample position caused by the Fe pin below.
This is particularly beneficial for adiabatic compression of the positron beam at low energies.
As shown in Figure \ref{fig:fwhm-vs-energy}, the field compression reduces the beam diameter from 8.3~mm (\ac{fwhm}) in the case of an Al pin to 5.0~mm with the Fe pin.


\begin{figure}[t!bp]
\centering
\includegraphics{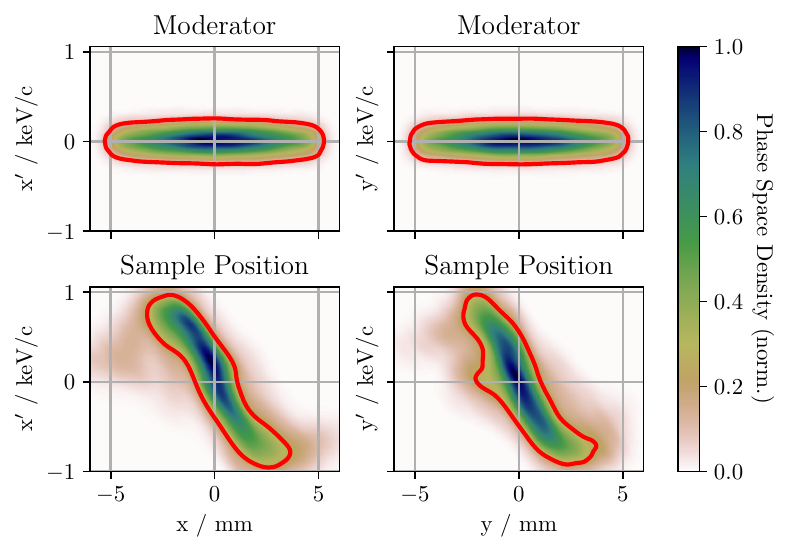}
\caption{\label{fig:phase-space-mod}
Transverse phase space volume occupied by positrons at the moderator (top) and the sample position (bottom) with 0~V acceleration, with the Fe pin below the sample holder.
The coordinates in real space and momentum space are denoted by x/y (as indicated in Figure~\ref{fig:sample-holder}) and $x^\prime$/$y^\prime$, respectively.
The red lines indicate 25\% of the maximum of each distribution.
At the moderator, the spatial distribution of positrons is described by a 4~mm circular, uniform source irradiating a flat surface at 4~mm distance, cut to the effective dimensions of the moderator ($10\times10~\text{mm}^2$).
We use 3~eV kinetic energy for the longitudinal momentum and a Maxwell-Boltzmann distribution with 40~meV mean thermal energy for the transverse
The focusing effect is seen as the distribution turns clockwise in both transverse directions.
That means that positrons to the left (right) of the center are moving to the right (left).
The focusing results from the increased magnetic flux density at the sample position due to the Fe pin below, as this adiabatically compresses the low-energy beam.
}
\end{figure}

\begin{figure}[h!tbp]
\centering
\includegraphics{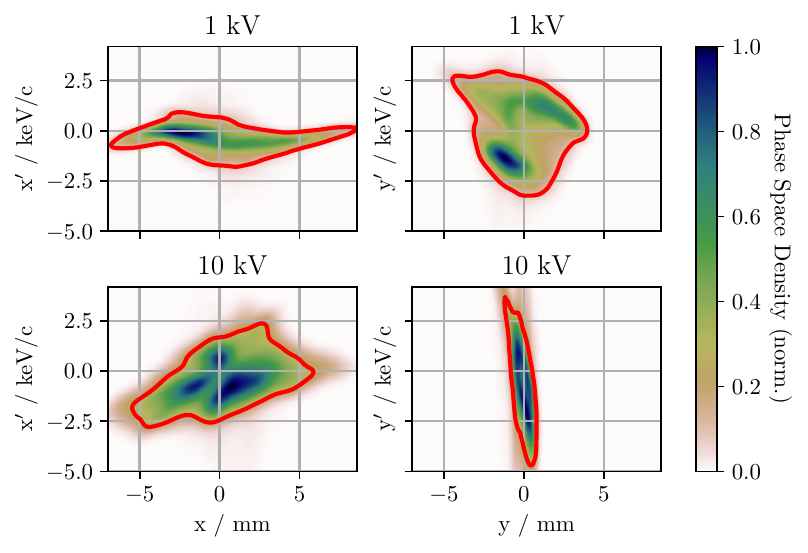}
\caption{\label{fig:phase-space}
Transverse phase space volume occupied by positrons at the sample position for an acceleration voltage of 1~kV (top) and 10~kV (bottom).
The coordinates in real space and momentum space are denoted by x/y (as indicated in Figure\,\ref{fig:sample-holder}) and $x^\prime$/$y^\prime$, respectively.
The red lines indicate 25\% of the maximum of each distribution.
The beam is more focused in both transverse directions ($x$ and $y$) at 10~kV compared to 1~kV, though the effect is much stronger in the $y$-direction, i.e., along the short side of the sample holder.
The momentum distributions broaden accordingly.
}
\end{figure}

Changes in the beam quality can be visualized by comparing the transverse phase space volume occupied by the monoenergetic positrons at the moderator and the sample position.
The phase space density is shown along the two transverse directions of the simulated beam of 50,000 positrons at the moderator and the sample position with no acceleration voltage (Figure \ref{fig:phase-space-mod}) and at the sample position for 1 and 10~kV acceleration voltage (Figure \ref{fig:phase-space}).
The phase space density is normalized to the maximum of all distributions, and the red lines indicate 25\% of the maximum of each distribution.

Figure \ref{fig:phase-space-mod} shows that the shape of the occupied phase space changes significantly when guiding the positrons to the sample position.
The beam diameter decreases from 5.8 to 5.0~mm (\ac{fwhm}) at the sample position.
As a result of Liouville's theorem, the distribution in transverse momentum-space ($x^\prime$ and $y^\prime$) gets broadened by the focusing.
The tilted shape ($\approx45^{\circ}$) of the occupied phase space at the sample position implies that positrons in the center of the beam have less transverse momentum than those at the periphery.
Specifically, and due to the Fe pin, the direction of the transverse momentum points back into the center.
The slight beginning of leveling of this tilt at large absolute values of $x$ and $y$ can be explained by the limited spatial extent of the magnetic field enhancement caused by the Fe pin.

The focusing force of the electric field at the sample holder dominates at implantation energies above $>5~$keV.
Since the long, rectangular sample holder with 20~mm width is positioned with its long sides parallel to the grounded sample chamber walls, the electric focusing is very strong along its short side ($y$ direction) and negligible along its long side ($x$ direction).
The comparison of different sample potentials in Figure \ref{fig:phase-space} shows the influence of the electrostatic focusing on the phase space distribution.
Note that the values of $x^\prime$ and $y^\prime$ are significantly larger than for the case of no acceleration voltage in Figure~\ref{fig:phase-space-mod}.
At 1~kV, we see only a very slight effect of the Fe pin (attributed to a $45^{\circ}$ tilt of the distribution reaching $\approx3~$mm in each direction), which vanishes completely for 10~kV.
For the 10~keV positron beam, the beam diameter, i.e., spread in space (x and y), gets smaller due to the larger electrostatic field.
The transverse momentum ($x^\prime$ and $y^\prime$) increases accordingly.
The large transverse momentum in the y-direction is due to the highly asymmetric electric field at the sample position caused by the grounded detector cups in the sample chamber.

\section{Measurements}\label{sec:measurements}

\subsection{Determining the Beam Diameter}\label{sec:beam-diameter}
\begin{figure}[htbp]
\centering
\includegraphics[width=\textwidth]{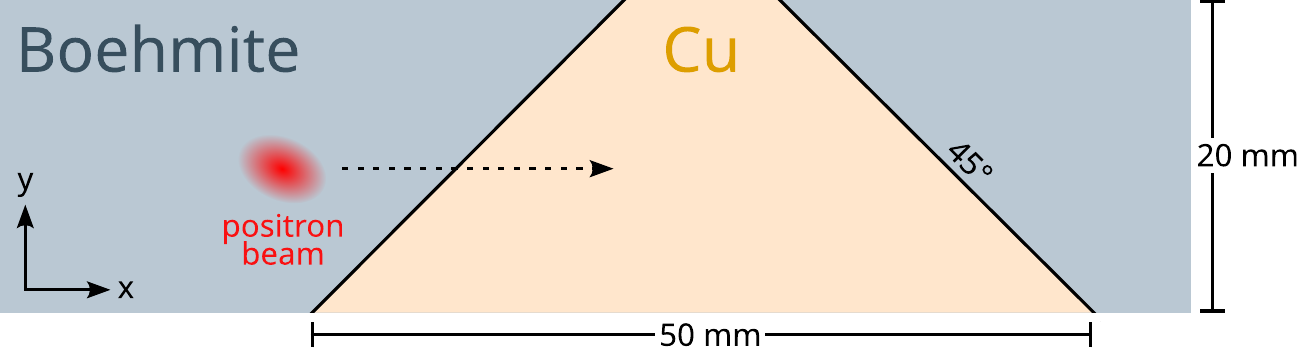}
\caption{\label{fig:sample-holder}
Schematic of the line scan sample holder consisting of a pseudoboehmite plate with Cu tape in a trapezoid shape on top.}
\end{figure}

The beam diameter can be determined using the knife-edge technique and line scans \cite{gigl2017}.
When dealing with higher positron energies, we typically utilize an Al/Cu edge as these materials have significantly different bulk S parameters.
However, for energies below 100~eV, the S values become similar as the line shape of the 511~keV photo peak is primarily affected by positron annihilation at the surface. 
Therefore, we rely on Ps formation measured through the \ac{v2p} ratio.
We have found that the combination of pseudoboehmite/Cu works well for both energy ranges.
Pseudoboehmite is a hydroxide oxide of aluminum, AlO(OH), with a slightly higher water content and less three-dimensional order than boehmite.
We produced a layer of pseudoboehmite (in the following called boehmite) on an Al sample holder by submerging it for one hour in boiling distilled water, which results in an amorphous and potentially porous layer with a thickness of $\approx350~$nm  \cite{altenpohl}.
The (open) porosity aids in the formation and escape of Ps, leading to a significant increase of \ac{v2p} at the surface.
To capture the beam shape in two independent directions using a linear feed-through, we use two 45° edges as shown in Figure~\ref{fig:sample-holder}.

\begin{figure}[htbp]
\centering
\includegraphics[width=\textwidth]{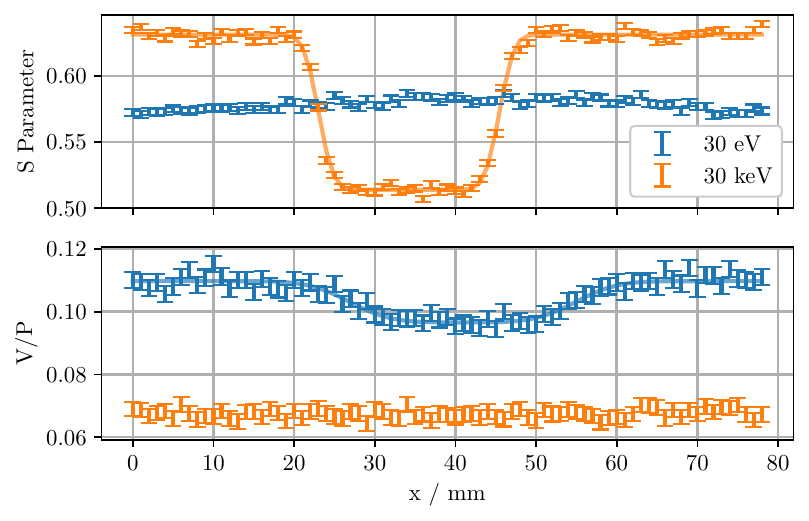}
\caption{\label{fig:line-scans} 
Line scans with the positron beam across two boehmite/Cu edges:
Comparison of S parameter and \ac{v2p} as a function of x at 33~eV and 30~keV implantation energy ($U_\text{mod}=30~$V).
The solid lines represent fits with error functions.}
\end{figure}

Figure \ref{fig:line-scans} presents line scans for 33 eV and 30 keV beam energies ($U_\text{mod}=30~$V).
We moved the sample holder in increments of 1 mm for a total of 78 mm.
Each measurement lasted 60~s and consists of $\approx1.5\cdot 10^5$ counts in the entire spectrum ($\approx4\cdot 10^4$ counts in the photo peak).
The plots show the position-dependent S parameter and \ac{v2p}.
It is worth noting that the S parameters of Cu and Al only show a clear difference at 30 keV but are nearly identical at 33 eV.
However, the \ac{v2p} data presents an opposing trend:
In the bulk, no Ps is formed at 30 keV positron implantation energy, leading to a constant \ac{v2p} independent of the material.
Near the surface, the different formation probability of Ps leads to a material-dependent change of \ac{v2p}.
(See section~\ref{sec:boehmite} for a complete depth profile measurement of boehmite.)
The data were fitted with error functions where possible to determine the beam diameter.
The according values are 10~mm (33~eV) and 3.1~mm (30~keV) \ac{fwhm}.

\begin{figure}[htbp]
\centering
\includegraphics[width=\textwidth]{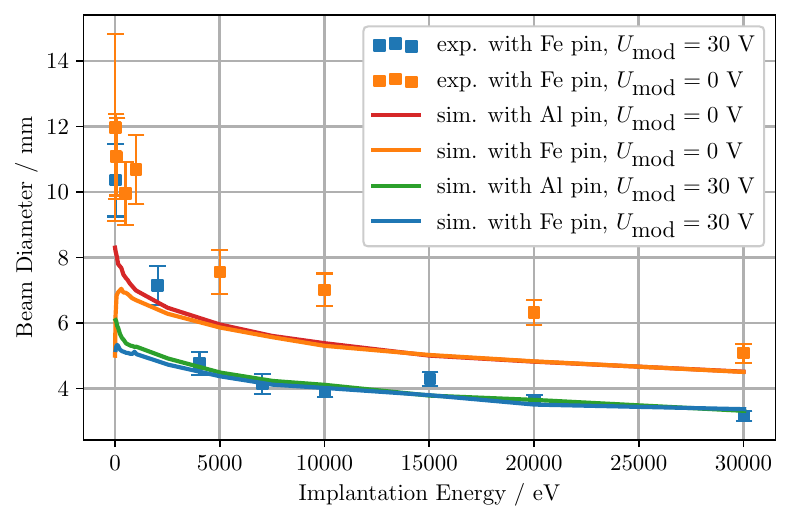}
\caption{\label{fig:fwhm-vs-energy}
Experimental data (squares) and simulation results (solid lines) of the energy-dependent beam diameter (\ac{fwhm}) at the sample position.
The simulation results show the benefit of using a magnetized Fe pin below the sample position for lower beam energies.
All data show the advantage (a significant decrease in beam diameter) of biasing the moderator to $U_\text{mod}=30~$V instead of 0~V.}
\end{figure}

Figure \ref{fig:fwhm-vs-energy} shows the beam diameter as a function of positron energy for both simulated and experimental data.
We performed simulations for 18 (20) different implantation energies for $U_\text{mod}=0~$V  ($30~$V) for a sample holder equipped with an Fe pin (blue and orange).
The beam diameters are calculated according to the knife-edge technique.
We repeated the simulations with an Al pin (green and red) to quantify the focusing effect of the magnetized Fe, particularly at low beam energies.
We used the Fe pin in all measurements presented in this publication.
The experimental data stems from fits to line scan data at each energy (as in Figure \ref{fig:line-scans}).
The lowest energy shown in the plot is 14~eV ($U_\text{mod}=0~$V) with a beam diameter of ($12\pm3$)~mm.
We observe an overall decreasing trend toward higher beam energies.
This behavior is expected due to the electrostatic focusing by the sample holder potential.
For the simulation with Fe pin and $U_\text{mod}=0~$V, we observe an additional decline towards low energies, causing a maximum diameter at $\approx300~$eV of 7.0~mm.
At 3~eV, the Fe pin improves the beam focus by 40\% (from 8.3~mm to 5.0~mm) compared to the Al pin.
Line scans recorded with $U_\text{mod}=0~$V (orange) roughly match the shape of the simulation data but exhibit an offset towards larger beam diameters.
The difference becomes even larger towards low energies, showing almost double the simulation values.
We believe this deviation is caused by the deterioration of the beam quality by external magnetic fields in the laboratory and imperfections of the magnetic field generating coils that are not considered in the simulation.
External magnetic fields can influence not only the beam direction (this is mostly compensated using the correction coils) but also the occupied phase space volume, increasing the beam diameter.
At only 3~eV kinetic energy, the positron beam is most prone to such effects.
The increasing deviations at low energies are due to difficulties in the line scan analysis caused by the diminishing contrast of \ac{v2p} for Cu and boehmite.

Experimental data at $U_\text{mod}=30~$V (blue) matches the simulation quite well at beam energies $>4~$keV.
Increasing the moderator bias facilitates beam guiding and significantly reduces the beam diameter at all acceleration voltages.
The two data points at lower energies differ significantly, displaying larger beam diameters.
We again explain this deviation with the low contrast of \ac{v2p} for Cu and boehmite.
In addition, the conservative assumptions regarding momentum and real space distribution at the moderator could partially explain the observed deviation.


\subsection{Ps Formation on Boehmite}\label{sec:boehmite}
\begin{figure}[htbp]
\centering
\includegraphics[width=\textwidth]{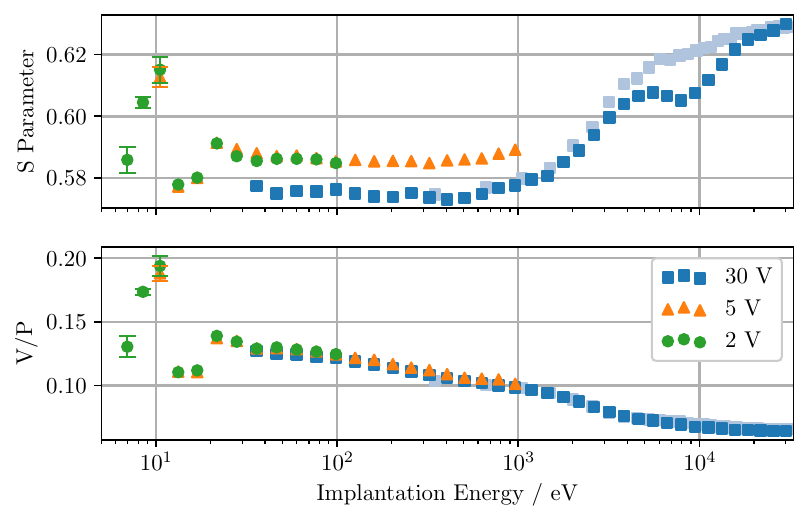}
\caption{\label{fig:boehmite-depth-profile}
Depth profiles S(E) and \ac{v2p}(E) of boehmite recorded at \ac{slope}.
The three colors correspond to different moderator potentials;
the light blue data stems from measurements of boehmite before N exposure.
The large peak observed for S and \ac{v2p} around 10~eV indicates a large fraction of produced free o-Ps.
}
\end{figure}

For boehmite on top of Al (as used in our line can sample holder in section \ref{sec:beam-diameter}), we recorded positron annihilation spectra as a function of positron implantation energy to obtain \ac{dbs} and \ac{v2p} depth profiles.
Figure \ref{fig:boehmite-depth-profile} shows the data covering an implantation energy range of 7~eV to 30~keV.
The blue markers represent data recorded with $U_\textrm{mod}=30$~V.
In this mode, we recorded the profile for both the pristine boehmite (light blue markers) and after exposure to nitrogen at 1~bar for 60~s (dark blue).
The other data (orange and green) show the low-energy settings of \ac{slope} with moderator voltages as low as 5 and 2~V, respectively.
The lowest sample potential in this measurement campaign was $U_\textrm{s}=-2$~V resulting in a minimum positron implantation energy of $E=-\Phi_+^\text{W} + U_\textrm{mod} - U_\textrm{s}$ = 7~eV.
Each data point is the average of ten S (\ac{v2p}) values, each from a 100~s measurement.
The typical photo peak count rate for beam energies above 5~keV is 1.2~kcts/s and slightly decreases towards lower kinetic energies.
Both plots (Figure \ref{fig:boehmite-depth-profile}) show the implantation energy in a logarithmic scale to clearly visualize the low-energy region.

The S(E) data for the 30~V setting shows a typical shape starting at a low surface S parameter and moving towards a high value of the bulk ($S_\textrm{bulk}$).
After N$_2$ exposure, a peak at $\approx6$~keV with a dip at slightly higher energy is observed. 
The reduced S parameter is attributed to a lower concentration of vacancy-like defects and pores (and possibly smaller in size) due to N$_2$ diffusion into boehmite. 
The minimum in the boehmite depth profile at 8~keV correlates to a mean positron implantation depth of $\approx300$~nm (using a density of 3.04 g/cm$^3$ for boehmite \cite{anthony}).
This corresponds to the depth of the boehmite-Al interface ($\approx350$~nm) and, thus, corroborates the interpretation that the deviation of the measured curve for pristine and N$_2$-loaded boehmite stems from filling vacancy-like defects with N$_2$.
Note that at energies higher than 20~keV, both curves coincide as is expected if positrons in the Al substrate dominate annihilation.


The low-energy S(E) profiles exhibit the same shape as the high-energy data while showing a constant offset towards higher S.
This well-known effect is caused by lower count rates, resulting in higher energy resolution and, hence, systematically higher S parameters.
Since the \ac{v2p} parameter is almost entirely independent of energy resolution, the according profiles show perfectly overlapping data points, confirming the excellent reproducibility of the measurements.
Its steady increase towards lower energies is related to positron back-diffusion and Ps formation at the surface.

The behavior of the data at the surface, i.e., at positron implantations energies lower than about 20~eV, and in particular, the large peak observed around 10~eV, clearly indicates the increased formation of free o-Ps.
First, the trend observed for S coincides with that of \ac{v2p}.
Two factors cause this effect:
Firstly, the increased formation of p-Ps annihilating with low momentum leads to a higher S parameter.
Secondly, the larger fraction of free o-Ps escaping to the vacuum and annihilating into three $\upgamma$ quanta (higher \ac{v2p}) is related to fewer annihilation events via the pick-off process (usually leading to larger Doppler-shifts), which also results in an increased S parameter.

The energy-dependent formation of o-Ps can be explained by accounting for the ionization energy of the material $E_\text{i}$ and the Ps binding energy $E_\text{Ps}=6.8$~eV.
For a quantitative estimation, we apply the Ore gap model \cite{humberston1986}, which describes the range of positron energies $E_+$ allowed for Ps formation:
$E_\text{i}-E_\text{Ps}<E_+<E_\text{i}$. 
Despite the low number of data points, we can estimate the ionization energy $E_\text{i}$ at the surface using two independent approaches. 
Assuming the peak energy is in the center of the Ore gap, we obtain 10 eV $=E_\text{i}-E_\text{Ps}/2$ and hence $E_\text{i}=13.4~$eV.
Alternatively, taking $E_+=7$~eV as the lower threshold for Ps formation results in  $E_\text{i}=E_++E_\text{Ps}=13.8~$eV.
Note that both values match very well with the ionization energy of H of 13.6~eV.
This agreement can be well explained by the hydrogen-terminated surface of boehmite, where the Ps is formed \cite{vedder1969}.

\subsection{Coincidence Depth Profile of Cu} 
\begin{figure}[htbp]
\centering
\includegraphics{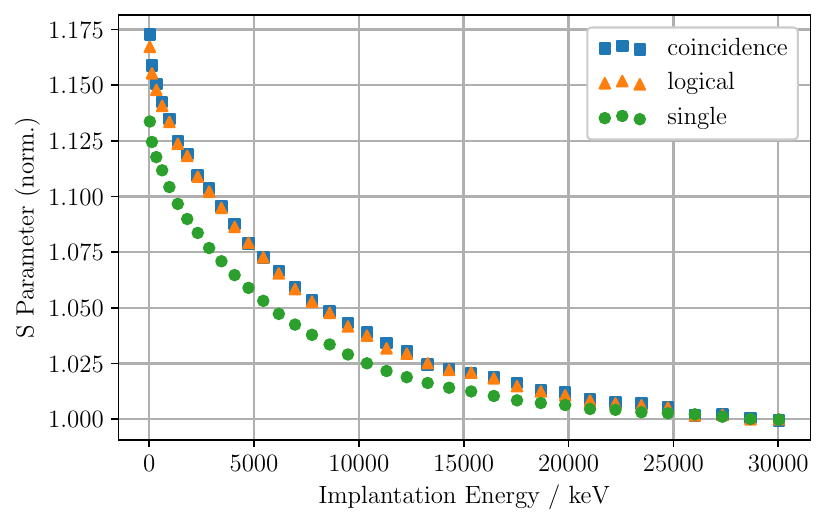}
\caption{\label{fig:copper-depth-profiles}
Depth profiles of Cu recorded with different measurement modes at \ac{slope}.
Conventional DBS ``single'', ``logical'' coincidence, and \ac{cdbs} ``coincidence'' exhibit different changes in the S parameter, i.e., different depths of information over the energy range covered (33~eV to 30~keV). The S parameter shown is normalized to the bulk value obtained for each measurement mode at the largest energy.}
\end{figure}

\ac{slope} allows conventional DBS and the measurement of \ac{v2p} as a function of positron implantation energy using HPGe detectors in single mode.
Using both detectors in coincidence, we can also perform depth-resolved CDBS, as demonstrated here for Cu.
We recorded annihilation spectra at 36 energies ranging from 33~eV to 30~keV.
Figure \ref{fig:copper-depth-profiles} illustrates the coincidence data acquisition capabilities of \ac{slope}.
Each data point consists of an average of at least eight 20~min measurements.
The three depth profiles shown in the plot are extracted from the same data using different processing methods:
``Single'' stands for the default S parameter calculations from the energy spectrum of one detector as it is usually performed for \ac{dbs}.
``Logical'' and ``coincidence'' look for coincident events (within 300~ns) in the time stamps of the two opposing detectors.
In addition to ``logical'', ``coincidence'' sets the condition to extract only coincident events with a combined energy of ($1022.0\pm0.5$)~keV, i.e., events in a diagonal region of interest of a 2D histogram containing only coincident events of two annihilation $\upgamma$ quanta detected in both detectors \cite{chryssos2023}.
The same overall trend, typical for Cu, is observed in all three depth profiles, ranging from a large surface ($S_\textrm{surf}$) to a lower bulk ($S_\textrm{bulk}$) value.
However, the total range of S values is clearly larger for the two coincidence methods, as expected from the improved background suppression. 
This results in a larger depth of information and makes smaller changes in the depth profiles visible.
Due to the reduced statistics, the error bars for the two coincidence modes are, while still in the order of the marker sizes, on average, five times larger (even slightly larger for ``coincidence'') than the single-mode error bars.


\section{Conclusion and Outlook}
The positron beam of \ac{slope} presented here was specifically designed for surface and near-surface studies with low-energy positrons.
Simulations conducted to understand beam characteristics and optimize coil currents highlight the effectiveness of the magnetic guidance system, particularly in achieving a stable beam position and minimum beam spot at the sample, independent of energy.
The comparison of simulations with experimental results not only validates the performance of \ac{slope} but also provides insights for further improvements.

We presented a new scheme to determine the position and diameter of the positron beam at very low positron energy.
To apply the knife-edge method in line scans across Cu/boehmite edges, we exploit the particularly large contrast of the \ac{v2p} values of these materials.
We demonstrate the versatility and performance of \ac{slope} by showing recorded depth profiles of boehmite to reveal valuable information about surface characteristics and Ps formation.
The observed Ps formation on hydrogen-terminated boehmite agrees well with theoretical considerations showcasing \ac{slope}'s potential as a setup dedicated to (near-) surface Ps spectroscopy.
In addition, different modes of data collection  by comparing \ac{cdbs} profiles of Cu have been presented.

With \ac{slope}, we provide a unique device that offers broad opportunities for monoenergetic positron beam studies of surfaces and near-surface regions of samples with enhanced precision and sensitivity.
Future research may include exploring new materials, functional surfaces, and complex interfaces and understanding the fundamental mechanisms governing Ps formation at and near surfaces.




\acknowledgments
Financial support by the German Research Foundation (DFG) within the Project HU 978/19-1 is gratefully acknowledged.


\end{document}

%% file: sample.tikzstyles
\tikzstyle{black dot}=[fill=black, draw=black, shape=circle]

\tikzstyle{black filled line}=[-, fill=black, thick, tikzit fill=black, tikzit draw=black]
\tikzstyle{light grey filled line}=[-, fill={rgb,255: red,186; green,186; blue,186}, tikzit fill={rgb,255: red,202; green,202; blue,202}, draw={rgb,255: red,186; green,186; blue,186}]
\tikzstyle{dark grey filled line}=[-, fill={rgb,255: red,104; green,104; blue,104}, line width=10pt, draw=none, tikzit fill={rgb,255: red,104; green,104; blue,104}, tikzit draw={rgb,255: red,126; green,126; blue,126}]
\tikzstyle{thick black line}=[-, line width=5pt]
\tikzstyle{dashed line}=[-, dashed]
\tikzstyle{fluid}=[-, fill=none, draw={rgb,255: red,216; green,216; blue,216}, line width=3pt, tikzit draw={rgb,255: red,216; green,216; blue,216}]
\tikzstyle{edgele}=[draw=black, fill=none, tikzit draw=black, {|-|}]
\tikzstyle{detector}=[-, fill={rgb,255: red,225; green,225; blue,225}, tikzit fill=white, draw=none, tikzit draw={rgb,255: red,226; green,226; blue,226}]
\tikzstyle{beamline}=[-, fill={rgb,255: red,218; green,218; blue,218}, tikzit fill={rgb,255: red,218; green,218; blue,218}]
\tikzstyle{fillblack}=[-, fill=black]
\tikzstyle{beamarrow}=[fill=white, draw={rgb,255: red,255; green,0; blue,4}, ->]
\tikzstyle{idiot}=[-, fill=white, draw=white]
\tikzstyle{arrowblack}=[tikzit draw=black, ->]
\tikzstyle{tube}=[-, thick, tikzit draw=black, fill={rgb,255: red,180; green,180; blue,180}, tikzit fill={rgb,255: red,180; green,180; blue,180}]
\tikzstyle{coffin}=[-, fill={rgb,255: red,89; green,89; blue,89}, tikzit fill={rgb,255: red,116; green,116; blue,116}, tikzit draw=black]
\tikzstyle{manipulator}=[-, fill={rgb,255: red,75; green,75; blue,75}, tikzit fill={rgb,255: red,75; green,75; blue,75}, tikzit draw=black]
\tikzstyle{manipulatearrow}=[tikzit draw=black, <->]
\tikzstyle{foil}=[-, line width=5pt, draw=black, tikzit draw=black]
\tikzstyle{ironpin}=[-, fill={rgb,255: red,157; green,157; blue,184}, tikzit fill={rgb,255: red,157; green,157; blue,184}, draw={rgb,255: red,157; green,157; blue,184}, tikzit draw={rgb,255: red,157; green,157; blue,184}]
\tikzstyle{coils}=[-, fill={rgb,255: red,35; green,35; blue,35}, tikzit fill={rgb,255: red,35; green,35; blue,35}, draw={rgb,255: red,35; green,35; blue,35}, tikzit draw={rgb,255: red,35; green,35; blue,35}]
\tikzstyle{tifoil}=[-, draw={rgb,255: red,126; green,42; blue,42}, tikzit draw={rgb,255: red,126; green,42; blue,42}, line width=2pt]
\tikzstyle{sampleholder}=[-, draw=black, fill=white, line width=1pt]

%% file: figures/labbeam.tikz
\begin{tikzpicture}[scale=0.4]
	\begin{pgfonlayer}{nodelayer}
		\node [style=none] (11) at (-18, 1) {};
		\node [style=none] (12) at (18, 1) {};
		\node [style=none] (13) at (18, -1) {};
		\node [style=none] (14) at (-18, -1) {};
		\node [style=none] (15) at (-18, -2) {};
		\node [style=none] (16) at (-17, -2) {};
		\node [style=none] (17) at (-17, -5) {};
		\node [style=none] (18) at (-14, -5) {};
		\node [style=none] (19) at (-14, -7) {};
		\node [style=none] (20) at (-17, -7) {};
		\node [style=none] (21) at (-17, -10) {};
		\node [style=none] (22) at (-16, -10) {};
		\node [style=none] (23) at (-16, -13) {};
		\node [style=none] (24) at (-19.5, -2) {};
		\node [style=none] (25) at (-20.5, -2) {};
		\node [style=none] (26) at (-20.5, -5) {};
		\node [style=none] (27) at (-20.5, -10) {};
		\node [style=none] (28) at (-21.5, -10) {};
		\node [style=none] (29) at (-21.5, -13) {};
		\node [style=none] (30) at (-19.5, -1) {};
		\node [style=none] (31) at (-21, -1) {};
		\node [style=none] (32) at (-21, 1) {};
		\node [style=none] (33) at (-21, 0.5) {};
		\node [style=none] (34) at (-19.25, 0.5) {};
		\node [style=none] (35) at (-19.25, -0.5) {};
		\node [style=none] (36) at (-21, -0.5) {};
		\node [style=none] (37) at (18, 0.75) {};
		\node [style=none] (38) at (18, -0.75) {};
		\node [style=none] (39) at (25.75, -6) {};
		\node [style=none] (40) at (24.25, -6) {};
		\node [style=none] (41) at (24.5, -7) {};
		\node [style=none] (42) at (25.5, -7) {};
		\node [style=none] (43) at (24.5, -7.5) {};
		\node [style=none] (44) at (24.5, -8) {};
		\node [style=none] (45) at (25.5, -7.5) {};
		\node [style=none] (46) at (25.5, -8) {};
		\node [style=none] (47) at (24.5, -8.5) {};
		\node [style=none] (48) at (25.5, -8.5) {};
		\node [style=none] (49) at (24.25, -7) {};
		\node [style=none] (50) at (25.75, -7) {};
		\node [style=none] (51) at (23.5, -8.5) {};
		\node [style=none] (55) at (26.5, -8.5) {};
		\node [style=none] (56) at (23.5, -10.5) {};
		\node [style=none] (57) at (26.5, -10.5) {};
		\node [style=none] (58) at (28.5, -10.5) {};
		\node [style=none] (59) at (21.5, -10.5) {};
		\node [style=none] (60) at (28.5, -13.75) {};
		\node [style=none] (61) at (21.5, -13.75) {};
		\node [style=none] (62) at (26.5, -15.75) {};
		\node [style=none] (63) at (23.5, -15.75) {};
		\node [style=none] (64) at (26.5, -13.75) {};
		\node [style=none] (65) at (23.5, -13.75) {};
		\node [style=none] (66) at (-18.75, -11) {Ion Getter};
		\node [style=none] (66) at (-18.75, -12) {Pump};
		\node [style=none] (67) at (-27.5, 1) {};
		\node [style=none] (68) at (-27.5, -1) {};
		\node [style=none] (69) at (-27.5, 2.75) {};
		\node [style=none] (70) at (-27.5, -2.75) {};
		\node [style=none] (71) at (-34, 2.75) {};
		\node [style=none] (72) at (-34, -2.75) {};
		\node [style=none] (73) at (-21, 1) {};
		\node [style=none] (74) at (-21, -1) {};
		\node [style=none] (75) at (-21, 0.5) {};
		\node [style=none] (76) at (-21, -0.5) {};
		\node [style=none] (77) at (-19.25, -0.5) {};
		\node [style=none] (78) at (-19.25, 0.5) {};
		\node [style=none] (79) at (-37, 0.15) {};
		\node [style=none] (80) at (-37, -0.15) {};
		\node [style=none] (81) at (-20.5, 0.35) {};
		\node [style=none] (82) at (-20.5, -0.35) {};
		\node [style=none] (83) at (-19.5, -0.35) {};
		\node [style=none] (84) at (-19.5, 0.35) {};
		\node [style=none] (85) at (-20.5, 0.15) {};
		\node [style=none] (86) at (-20.5, -0.15) {};
		\node [style=none] (87) at (-19.5, 1) {};
		\node [style=none] (88) at (-18, 1) {};
		\node [style=none] (89) at (-19.5, 2) {};
		\node [style=none] (90) at (-18, 2) {};
		\node [style=none] (91) at (-19.875, 2) {};
		\node [style=none] (92) at (-17.625, 2) {};
		\node [style=none] (93) at (-19.875, 2.5) {};
		\node [style=none] (94) at (-17.625, 2.5) {};
		\node [style=none] (95) at (-18.25, 2.55) {};
		\node [style=none] (96) at (-19.25, 2.55) {};
		\node [style=none] (97) at (-19.25, 2.75) {};
		\node [style=none] (98) at (-18.25, 2.75) {};
		\node [style=none] (99) at (-18.9, 2.75) {};
		\node [style=none] (100) at (-18.6, 2.75) {};
		\node [style=none] (101) at (-18.9, 4) {};
		\node [style=none] (102) at (-18.6, 4) {};
		\node [style=none] (103) at (-18, 3.5) {};
		\node [style=none] (104) at (-18, 4.75) {};
		\node [style=none] (105) at (-35.25, 0.75) {};
		\node [style=none] (106) at (-36.5, 0.75) {};
		\node [style=none] (108) at (-18.75, -1) {};
		\node [style=none] (109) at (-18.75, 1) {};
		\node [style=none] (110) at (-15.5, -6) {};
		\node [style=none] (111) at (-15.75, -6) {TMP 1};
		\node [style=none] (112) at (-6.25, 1) {};
		\node [style=none] (113) at (-5.75, 1) {};
		\node [style=none] (114) at (-6, 0) {};
		\node [style=none] (115) at (-5.75, -1) {};
		\node [style=none] (116) at (-6.25, -1) {};
		\node [style=none] (117) at (-6.25, -2) {};
		\node [style=none] (118) at (-5.75, -2) {};
		\node [style=none] (119) at (-6.25, 4) {};
		\node [style=none] (120) at (-5.75, 4) {};
		\node [style=none] (121) at (-6, 4.75) {Gate Valve S1};
		\node [style=none] (122) at (17, 1) {};
		\node [style=none] (123) at (17.5, 1) {};
		\node [style=none] (124) at (17.25, 0) {};
		\node [style=none] (125) at (17.5, -1) {};
		\node [style=none] (126) at (17, -1) {};
		\node [style=none] (127) at (17, -2) {};
		\node [style=none] (128) at (17.5, -2) {};
		\node [style=none] (129) at (17, 4) {};
		\node [style=none] (130) at (17.5, 4) {};
		\node [style=none] (131) at (17.25, 4.75) {Optional Aperture};
		\node [style=none] (132) at (26, -6) {};
		\node [style=none] (133) at (26, -6.5) {};
		\node [style=none] (134) at (25, -6.25) {};
		\node [style=none] (135) at (24, -6.5) {};
		\node [style=none] (136) at (24, -6) {};
		\node [style=none] (137) at (23, -6) {};
		\node [style=none] (138) at (23, -6.5) {};
		\node [style=none] (139) at (29, -6) {};
		\node [style=none] (140) at (29, -6.5) {};
		\node [style=none] (141) at (29.75, -5) {Gate Valve S2};
		\node [style=none] (142) at (-13, 0) {G1};
		\node [style=none] (143) at (6.25, 1.35) {};
		\node [style=none] (144) at (5.75, 1.35) {};
		\node [style=none] (145) at (5.75, -1.375) {};
		\node [style=none] (146) at (6.25, -1.375) {};
		\node [style=none] (147) at (5.75, -1) {};
		\node [style=none] (148) at (6.25, -1) {};
		\node [style=none] (149) at (5.75, 1) {};
		\node [style=none] (150) at (6.25, 1) {};
		\node [style=none] (151) at (-0.5, 0) {G2};
		\node [style=none] (152) at (11.25, 0) {G3};
		\node [style=none] (153) at (28.5, -13.5) {};
		\node [style=none] (154) at (25.875, -13.5) {};
		\node [style=none] (155) at (28.5, -10.75) {};
		\node [style=none] (156) at (25.875, -10.75) {};
		\node [style=none] (157) at (24.125, -13.5) {};
		\node [style=none] (158) at (21.5, -13.5) {};
		\node [style=none] (159) at (24.125, -10.75) {};
		\node [style=none] (160) at (21.5, -10.75) {};
		\node [style=none] (161) at (22.8, -12.125) {Det 1};
		\node [style=none] (162) at (27.25, -12.125) {Det 2};
		\node [style=none] (163) at (24.625, -15.7) {};
		\node [style=none] (164) at (25.375, -15.7) {};
		\node [style=none] (165) at (24.625, -13.25) {};
		\node [style=none] (167) at (24.75, -13) {};
		\node [style=none] (168) at (25.375, -13.25) {};
		\node [style=none] (169) at (25.25, -13) {};
		\node [style=none] (170) at (24.925, -12.95) {};
		\node [style=none] (171) at (25.075, -12.95) {};
		\node [style=none] (172) at (24.925, -12.45) {};
		\node [style=none] (173) at (25.075, -12.45) {};
		\node [style=none] (174) at (24.75, -12.4) {};
		\node [style=none] (175) at (25.25, -12.4) {};
		\node [style=none] (176) at (26.75, -10.25) {};
		\node [style=none] (177) at (26.75, -9.75) {};
		\node [style=none] (178) at (27.25, -9.75) {};
		\node [style=none] (179) at (27.25, -10.25) {};
		\node [style=none] (180) at (26.75, -9.25) {};
		\node [style=none] (181) at (26.75, -8.75) {};
		\node [style=none] (182) at (27.25, -8.75) {};
		\node [style=none] (183) at (27.25, -9.25) {};
		\node [style=none] (184) at (26.75, -8.5) {};
		\node [style=none] (185) at (26.75, -8) {};
		\node [style=none] (186) at (27.25, -8) {};
		\node [style=none] (187) at (27.25, -8.5) {};
		\node [style=none] (188) at (22.75, -8.5) {};
		\node [style=none] (189) at (22.75, -8) {};
		\node [style=none] (190) at (23.25, -8) {};
		\node [style=none] (191) at (23.25, -8.5) {};
		\node [style=none] (192) at (22.75, -9.25) {};
		\node [style=none] (193) at (22.75, -8.75) {};
		\node [style=none] (194) at (23.25, -8.75) {};
		\node [style=none] (195) at (23.25, -9.25) {};
		\node [style=none] (196) at (22.75, -10.25) {};
		\node [style=none] (197) at (22.75, -9.75) {};
		\node [style=none] (198) at (23.25, -9.75) {};
		\node [style=none] (199) at (23.25, -10.25) {};
		\node [style=none] (200) at (17.5, -14.7) {Sample Holder};
		\node [style=none] (201) at (18.75, -14) {};
		\node [style=none] (202) at (25.5, -12.5) {};
		\node [style=none] (203) at (25.5, -12.5) {};
		\node [style=none] (204) at (24.5, -12.5) {};
		\node [style=none] (205) at (29.6, -8.2) {TMP 2};
		\node [style=none] (206) at (28.75, -8.75) {};
		\node [style=none] (207) at (28, -9.75) {};
		\node [style=none] (208) at (22, -9.5) {H2};
		\node [style=none] (209) at (22, -8.25) {F9};
		\node [style=none] (210) at (23.5, -7.25) {};
		\node [style=none] (211) at (23.5, -6.75) {};
		\node [style=none] (212) at (24, -6.75) {};
		\node [style=none] (213) at (24, -7.25) {};
		\node [style=none] (214) at (23.5, -5.75) {};
		\node [style=none] (215) at (23.5, -5.25) {};
		\node [style=none] (216) at (24, -5.25) {};
		\node [style=none] (217) at (24, -5.75) {};
		\node [style=none] (218) at (26, -7.25) {};
		\node [style=none] (219) at (26, -6.75) {};
		\node [style=none] (220) at (26.5, -6.75) {};
		\node [style=none] (221) at (26.5, -7.25) {};
		\node [style=none] (222) at (26, -5.75) {};
		\node [style=none] (223) at (26, -5.25) {};
		\node [style=none] (224) at (26.5, -5.25) {};
		\node [style=none] (225) at (26.5, -5.75) {};
		\node [style=none] (226) at (22.5, -5.5) {F7};
		\node [style=none] (227) at (22.5, -7) {F8};
		\node [style=none] (228) at (23.25, -1.3) {G4};
		\node [style=none] (229) at (17.75, -1.75) {};
		\node [style=none] (230) at (17.75, -1.25) {};
		\node [style=none] (231) at (18.25, -1.25) {};
		\node [style=none] (232) at (18.25, -1.75) {};
		\node [style=none] (233) at (16.25, -1.75) {};
		\node [style=none] (234) at (16.25, -1.25) {};
		\node [style=none] (235) at (16.75, -1.25) {};
		\node [style=none] (236) at (16.75, -1.75) {};
		\node [style=none] (237) at (17.75, 1.25) {};
		\node [style=none] (238) at (17.75, 1.75) {};
		\node [style=none] (239) at (18.25, 1.75) {};
		\node [style=none] (240) at (18.25, 1.25) {};
		\node [style=none] (241) at (16.25, 1.25) {};
		\node [style=none] (242) at (16.25, 1.75) {};
		\node [style=none] (243) at (16.75, 1.75) {};
		\node [style=none] (244) at (16.75, 1.25) {};
		\node [style=none] (245) at (16.5, -2.75) {F5};
		\node [style=none] (246) at (18, -2.75) {F6};
		\node [style=none] (247) at (-0.5, 2) {C2-H/V};
		\node [style=none] (248) at (11.25, 2) {C3-H/V};
		\node [style=none] (249) at (-13, 2) {C1-H/V};
		\node [style=none] (250) at (-5.5, 1.25) {};
		\node [style=none] (251) at (-5.5, 1.75) {};
		\node [style=none] (252) at (-5, 1.75) {};
		\node [style=none] (253) at (-5, 1.25) {};
		\node [style=none] (254) at (-7, 1.25) {};
		\node [style=none] (255) at (-7, 1.75) {};
		\node [style=none] (256) at (-6.5, 1.75) {};
		\node [style=none] (257) at (-6.5, 1.25) {};
		\node [style=none] (258) at (-5.5, -1.75) {};
		\node [style=none] (259) at (-5.5, -1.25) {};
		\node [style=none] (260) at (-5, -1.25) {};
		\node [style=none] (261) at (-5, -1.75) {};
		\node [style=none] (262) at (-7, -1.75) {};
		\node [style=none] (263) at (-7, -1.25) {};
		\node [style=none] (264) at (-6.5, -1.25) {};
		\node [style=none] (265) at (-6.5, -1.75) {};
		\node [style=none] (266) at (-6.75, -2.75) {F2};
		\node [style=none] (267) at (-5.25, -2.75) {F3};
		\node [style=none] (268) at (-20, 0) {};
		\node [style=none] (269) at (-24.25, -3.25) {};
		\node [style=none] (270) at (-19.25, 0) {};
		\node [style=none] (271) at (-15.75, -2.75) {};
		\node [style=none] (272) at (-18.625, 0) {};
		\node [style=none] (274) at (-24.25, -4) {$\upbeta^+$-Emitter};
		\node [style=none] (285) at (-25.25, 1) {};
		\node [style=none] (286) at (-24.75, 1) {};
		\node [style=none] (287) at (-25, 0) {};
		\node [style=none] (288) at (-24.75, -1) {};
		\node [style=none] (289) at (-25.25, -1) {};
		\node [style=none] (290) at (-25.25, -2) {};
		\node [style=none] (291) at (-24.75, -2) {};
		\node [style=none] (292) at (-25.25, 4) {};
		\node [style=none] (293) at (-24.75, 4) {};
		\node [style=none] (294) at (-25, 4.75) {Gate Valve S0};
		\node [style=none] (295) at (-14.25, -3) {Al Foil};
		\node [style=none] (296) at (-17.75, 1.25) {};
		\node [style=none] (297) at (-17.75, 1.75) {};
		\node [style=none] (298) at (-17.25, 1.75) {};
		\node [style=none] (299) at (-17.25, 1.25) {};
		\node [style=none] (300) at (-17.75, -1.25) {};
		\node [style=none] (301) at (-17.75, -1.75) {};
		\node [style=none] (302) at (-17.25, -1.25) {};
		\node [style=none] (303) at (-17.25, -1.75) {};
		\node [style=none] (304) at (-19.75, -1.25) {};
		\node [style=none] (305) at (-19.75, -1.75) {};
		\node [style=none] (306) at (-20.25, -1.75) {};
		\node [style=none] (307) at (-20.25, -1.25) {};
		\node [style=none] (308) at (-19.75, 1.25) {};
		\node [style=none] (309) at (-19.75, 1.75) {};
		\node [style=none] (310) at (-20.25, 1.75) {};
		\node [style=none] (311) at (-20.25, 1.25) {};
		\node [style=none] (312) at (-21, 1.5) {H1};
		\node [style=none] (313) at (-16.5, 3.5) {};
		\node [style=none] (314) at (-17, 1.75) {};
		\node [style=none] (315) at (-17, 1.25) {};
		\node [style=none] (316) at (-14.75, 1.25) {};
		\node [style=none] (317) at (-14.75, 1.75) {};
		\node [style=none] (318) at (-17, -1.25) {};
		\node [style=none] (319) at (-17, -1.75) {};
		\node [style=none] (320) at (-14.75, -1.75) {};
		\node [style=none] (321) at (-14.75, -1.25) {};
		\node [style=none] (322) at (-16, 2.5) {F1};
		\node [style=none] (323) at (-15, 4.25) {W Moderator};
		\node [style=none] (324) at (25.25, 0.25) {C4-H/V};
		\node [style=none] (325) at (5.75, 1.625) {};
		\node [style=none] (326) at (5.75, 2.125) {};
		\node [style=none] (327) at (6.25, 2.125) {};
		\node [style=none] (328) at (6.25, 1.625) {};
		\node [style=none] (329) at (5.75, -2.125) {};
		\node [style=none] (330) at (5.75, -1.625) {};
		\node [style=none] (331) at (6.25, -1.625) {};
		\node [style=none] (332) at (6.25, -2.125) {};
		\node [style=none] (333) at (6, -2.75) {F4};
		\node [style=none] (334) at (26.75, -15.5) {};
		\node [style=none] (335) at (26.75, -15) {};
		\node [style=none] (336) at (27.25, -15) {};
		\node [style=none] (337) at (27.25, -15.5) {};
		\node [style=none] (338) at (26.75, -14.5) {};
		\node [style=none] (339) at (26.75, -14) {};
		\node [style=none] (340) at (27.25, -14) {};
		\node [style=none] (341) at (27.25, -14.5) {};
		\node [style=none] (342) at (22.75, -15.5) {};
		\node [style=none] (343) at (22.75, -15) {};
		\node [style=none] (344) at (23.25, -15) {};
		\node [style=none] (345) at (23.25, -15.5) {};
		\node [style=none] (346) at (22.75, -14.5) {};
		\node [style=none] (347) at (22.75, -14) {};
		\node [style=none] (348) at (23.25, -14) {};
		\node [style=none] (349) at (23.25, -14.5) {};
		\node [style=none] (350) at (28.25, -14.75) {H3};
		\node [style=none] (351) at (-30.75, -0.9) {\textcolor{white}{Source}};
		\node [style=none] (351) at (-30.75, -1.9) {\textcolor{white}{Container}};
		\node [style=none] (352) at (29, -9) {};
		\node [style=none] (353) at (-18.3, 0.5) {};
		\node [style=none] (354) at (-17.6, 0.5) {};
		\node [style=none] (355) at (-17.6, 0.7) {};
		\node [style=none] (356) at (-18.3, -0.5) {};
		\node [style=none] (357) at (-17.6, -0.5) {};
		\node [style=none] (358) at (-17.6, -0.7) {};
		\node [style=none] (359) at (-17.5, 0.7) {};
		\node [style=none] (360) at (-16.8, 0.7) {};
		\node [style=none] (361) at (-17.5, -0.7) {};
		\node [style=none] (362) at (-16.8, -0.7) {};
		\node [style=none] (363) at (-16.7, 0.7) {};
		\node [style=none] (364) at (-15.0, 0.7) {};
		\node [style=none] (365) at (-16.7, -0.7) {};
		\node [style=none] (366) at (-15.0, -0.7) {};
		\node [style=none] (367) at (-18.0, 0) {\tiny LA};
		\node [style=none] (368) at (-17.1, 0) {\tiny LB};
		\node [style=none] (368) at (-15.85, 0) {\tiny LC};
	\end{pgfonlayer}
	\begin{pgfonlayer}{edgelayer}
		\draw [style=tube] (34.center)
			 to (33.center)
			 to (32.center)
			 to (11.center)
			 to (12.center)
			 to (37.center)
			 to [in=90, out=0, looseness=1.25] (39.center)
			 to (50.center)
			 to (42.center)
			 to [bend right=90, looseness=1.25] (45.center)
			 to [bend right=90, looseness=1.25] (46.center)
			 to [bend right=90, looseness=1.50] (48.center)
			 to (55.center)
			 to (57.center)
			 to (58.center)
			 to (60.center)
			 to (64.center)
			 to (62.center)
			 to (63.center)
			 to (65.center)
			 to (61.center)
			 to (59.center)
			 to (56.center)
			 to (51.center)
			 to (47.center)
			 to [bend right=90, looseness=1.25] (44.center)
			 to [bend right=90, looseness=1.25] (43.center)
			 to [bend right=90, looseness=1.25] (41.center)
			 to (49.center)
			 to (40.center)
			 to [in=0, out=90, looseness=1.25] (38.center)
			 to (13.center)
			 to (14.center)
			 to (15.center)
			 to (16.center)
			 to (17.center)
			 to (18.center)
			 to (19.center)
			 to (20.center)
			 to (21.center)
			 to (22.center)
			 to (23.center)
			 to (29.center)
			 to (28.center)
			 to (27.center)
			 to (26.center)
			 to (25.center)
			 to (24.center)
			 to (30.center)
			 to (31.center)
			 to (36.center)
			 to (35.center)
			 to cycle;
		\draw [style=coffin] (74.center)
			 to (68.center)
			 to (70.center)
			 to (72.center)
			 to (71.center)
			 to (69.center)
			 to (67.center)
			 to (73.center)
			 to (75.center)
			 to (78.center)
			 to (77.center)
			 to (76.center)
			 to cycle;
		\draw [style=fillblack] (84.center)
			 to (81.center)
			 to (82.center)
			 to (83.center)
			 to cycle;
		\draw [style=tube] (85.center)
			 to (86.center)
			 to (80.center)
			 to (79.center)
			 to cycle;
        \draw (353.center) 
			 to (354.center)
			 to (355.center);
        \draw (356.center)
			 to (357.center)
          to (358.center);
        \draw (359.center) 
			 to (360.center);
        \draw (361.center)
          to (362.center);
        \draw (363.center) 
			 to (364.center);
        \draw (365.center)
          to (366.center);
		\draw [style=tube] (90.center)
			 to (89.center)
			 to (87.center)
			 to (88.center)
			 to cycle;
		\draw [style=tube] (93.center)
			 to (91.center)
			 to (92.center)
			 to (94.center)
			 to cycle;
		\draw [style=coffin] (96.center)
			 to (97.center)
			 to (98.center)
			 to (95.center)
			 to cycle;
		\draw [style=manipulator] (99.center)
			 to (100.center)
			 to (102.center)
			 to [in=360, out=180] (101.center)
			 to cycle;
		\draw [style=manipulatearrow] (104.center) to (103.center);
		\draw [style=manipulatearrow] (106.center) to (105.center);
		\draw [style=foil] (109.center) to (108.center);
		\draw [style=black filled line] (115.center)
			 to (114.center)
			 to (113.center)
			 to (120.center)
			 to (119.center)
			 to (112.center)
			 to (114.center)
			 to (116.center)
			 to (117.center)
			 to (118.center)
			 to cycle;
		\draw [style=black filled line] (125.center)
			 to (124.center)
			 to (123.center)
			 to (130.center)
			 to (129.center)
			 to (122.center)
			 to (124.center)
			 to (126.center)
			 to (127.center)
			 to (128.center)
			 to cycle;
		\draw [style=black filled line] (135.center)
			 to (134.center)
			 to (133.center)
			 to (140.center)
			 to (139.center)
			 to (132.center)
			 to (134.center)
			 to (136.center)
			 to (137.center)
			 to (138.center)
			 to cycle;
		\draw [style=dashed line] (149.center) to (147.center);
		\draw [style=dashed line] (148.center) to (150.center);
		\draw [style=manipulator] (143.center)
			 to (144.center)
			 to (149.center)
			 to (150.center)
			 to cycle;
		\draw [style=manipulator] (146.center)
			 to (148.center)
			 to (147.center)
			 to (145.center)
			 to cycle;
		\draw [style=detector] (153.center)
			 to [in=270, out=90] (155.center)
			 to [in=360, out=180] (156.center)
			 to [in=90, out=-90] (154.center)
			 to [in=180, out=0] cycle;
		\draw [style=detector] (157.center)
			 to [in=270, out=90] (159.center)
			 to [in=360, out=180] (160.center)
			 to [in=90, out=-90] (158.center)
			 to [in=180, out=0] cycle;
		\draw [style=dark grey filled line] (169.center)
			 to [bend left] (168.center)
			 to (164.center)
			 to (163.center)
			 to (165.center)
			 to [bend left] (167.center)
			 to cycle;
		\draw [style=ironpin] (171.center)
			 to (170.center)
			 to (172.center)
			 to (173.center)
			 to cycle;
		\draw [style=coils] (177.center)
			 to (176.center)
			 to (179.center)
			 to (178.center)
			 to cycle;
		\draw [style=coils] (181.center)
			 to (180.center)
			 to (183.center)
			 to (182.center)
			 to cycle;
		\draw [style=coils] (185.center)
			 to (184.center)
			 to (187.center)
			 to (186.center)
			 to cycle;
		\draw [style=coils] (189.center)
			 to (188.center)
			 to (191.center)
			 to (190.center)
			 to cycle;
		\draw [style=coils] (193.center)
			 to (192.center)
			 to (195.center)
			 to (194.center)
			 to cycle;
		\draw [style=coils] (197.center)
			 to (196.center)
			 to (199.center)
			 to (198.center)
			 to cycle;
		\draw [style=dashed line] (204.center) to (201.center);
		\draw [style=coils] (211.center)
			 to (210.center)
			 to (213.center)
			 to (212.center)
			 to cycle;
		\draw [style=coils] (215.center)
			 to (214.center)
			 to (217.center)
			 to (216.center)
			 to cycle;
		\draw [style=coils] (219.center)
			 to (218.center)
			 to (221.center)
			 to (220.center)
			 to cycle;
		\draw [style=coils] (223.center)
			 to (222.center)
			 to (225.center)
			 to (224.center)
			 to cycle;
		\draw [style=coils] (230.center)
			 to (229.center)
			 to (232.center)
			 to (231.center)
			 to cycle;
		\draw [style=coils] (234.center)
			 to (233.center)
			 to (236.center)
			 to (235.center)
			 to cycle;
		\draw [style=coils] (237.center)
			 to (240.center)
			 to (239.center)
			 to (238.center)
			 to cycle;
		\draw [style=coils] (243.center)
			 to (242.center)
			 to (241.center)
			 to (244.center)
			 to cycle;
		\draw [style=coils] (252.center)
			 to (251.center)
			 to (250.center)
			 to (253.center)
			 to cycle;
		\draw [style=coils] (256.center)
			 to (255.center)
			 to (254.center)
			 to (257.center)
			 to cycle;
		\draw [style=coils] (259.center)
			 to (258.center)
			 to (261.center)
			 to (260.center)
			 to cycle;
		\draw [style=coils] (263.center)
			 to (262.center)
			 to (265.center)
			 to (264.center)
			 to cycle;
		\draw [style=tifoil] (78.center) to (77.center);
		\draw [style=dashed line] (268.center) to (269.center);
		\draw [style=black filled line] (288.center)
			 to (287.center)
			 to (286.center)
			 to (293.center)
			 to (292.center)
			 to (285.center)
			 to (287.center)
			 to (289.center)
			 to (290.center)
			 to (291.center)
			 to cycle;
		\draw [style=coils] (298.center)
			 to (297.center)
			 to (296.center)
			 to (299.center)
			 to cycle;
		\draw [style=coils] (309.center)
			 to (308.center)
			 to (311.center)
			 to (310.center)
			 to cycle;
		\draw [style=coils] (306.center)
			 to (305.center)
			 to [in=270, out=90] (304.center)
			 to (307.center)
			 to cycle;
		\draw [style=coils] (303.center)
			 to [in=270, out=90] (302.center)
			 to (300.center)
			 to (301.center)
			 to cycle;
		\draw [style=coils] (319.center)
			 to (320.center)
			 to (321.center)
			 to (318.center)
			 to cycle;
		\draw [style=coils] (316.center)
			 to (317.center)
			 to (314.center)
			 to (315.center)
			 to cycle;
		\draw [style=dashed line] (313.center) to (272.center);
		\draw [style=dashed line] (270.center) to (271.center);
		\draw [style=coils] (325.center)
			 to (328.center)
			 to (327.center)
			 to (326.center)
			 to cycle;
		\draw [style=coils] (329.center)
			 to (332.center)
			 to (331.center)
			 to (330.center)
			 to cycle;
		\draw [style=coils] (335.center)
			 to (334.center)
			 to (337.center)
			 to (336.center)
			 to cycle;
		\draw [style=coils] (338.center)
			 to (341.center)
			 to (340.center)
			 to (339.center)
			 to cycle;
		\draw [style=coils] (344.center)
			 to (343.center)
			 to (342.center)
			 to (345.center)
			 to cycle;
		\draw [style=coils] (346.center)
			 to (349.center)
			 to (348.center)
			 to (347.center)
			 to cycle;
		\draw [style=black filled line] (207.center) to (206.center);
		\draw [style=black filled line] (207.center) to (352.center);
		\draw [style=black filled line] (352.center) to (206.center);
		\draw [style=sampleholder] (175.center) to (174.center);
	\end{pgfonlayer}
\end{tikzpicture}